# JACOBIAN MATRIX: A BRIDGE BETWEEN LINEAR AND NONLNEAR POLYNOMIAL-ONLY PROBLEMS

W. Chen

Permanent mail address: P. O. Box 2-19-201, Jiangshu University of Science & Technology, Zhenjiang City, Jiangsu Province 212013, P. R. China

Present mail address (as a JSPS Postdoctoral Research Fellow): Apt.4, West 1$^{st}$ floor, Himawari-so, 316-2, Wakasato-kitaichi, Nagano-city, Nagano-ken, 380-0926, JAPAN

E-mail: chenw@homer.shinshu-u.ac.jp

Permanent email box: chenwwhy@hotmail.com

Abbreviated Title: **Nonlinear linear polynomial-only problems**

**Abstract**: By using the Hadamard matrix product concept, this paper introduces two generalized matrix formulation forms of numerical analogue of nonlinear differential operators. The SJT matrix-vector product approach is found to be a simple, efficient and accurate technique in the calculation of the Jacobian matrix of the nonlinear discretization by finite difference, finite volume, collocation, dual reciprocity BEM or radial functions based numerical methods. We also present and prove simple underlying relationship (theorem (3.1)) between general nonlinear analogue polynomials and their corresponding Jacobian matrices, which forms the basis of this paper. By means of theorem 3.1, stability analysis of numerical solutions of nonlinear initial value problems can be easily handled based on the well-known results for linear problems. Theorem 3.1 also leads naturally to the straightforward extension of various linear iterative algorithms such as the SOR, Gauss-Seidel and Jacobi methods to nonlinear algebraic equations. Since an exact alternative of the quasi-Newton equation is established via theorem 3.1, we derive a modified BFGS quasi-Newton method. A simple formula is also given to examine the deviation between the approximate and exact Jacobian matrices. Furthermore, in order to avoid the evaluation of the Jacobian matrix and its inverse, the pseudo-Jacobian matrix is introduced with a general applicability of any nonlinear systems of equations. It should be pointed out that a large class of real-world nonlinear problems can be modeled or numerically discretized polynomial-only algebraic system of equations. The results presented here are in general applicable for all these problems. This paper can be considered as a starting point in the research of nonlinear computation and analysis from an innovative viewpoint.

**Key words**. Hadamard product, Jacobian matrix, SJT product, nonlinear polynomial-only equations, nonlinear stability analysis, quasi-Newton method, pseudo-Jacobian matrix.

**AMS subject classifications.** 47H17, 65J15

**1. Introduction.** The numerical solution of nonlinear partial differential equations plays a prominent role in many areas of physical and engineering. Considerable research endeavors have been directed to develop various nonlinear solution methodologies. However, it is not easy to achieve some significant practical progress in this direction because of great complexity the nonlinearity arises. Recently, some innovative contributions were made by one of the present authors [1, 2]. The Hadamard product of matrices was therein introduced to nonlinear computations and analysis. The SJT product of matrix and vector was defined to efficiently calculate the accurate Jacobian matrix of some numerical formulations of nonlinear differential equations. The present study is a step forward development based on these works.

In comparison to nonlinear cases, a vast variety of computing and analysis tools of linear problems have been quite well developed today. It is natural desire to employ these effective linear methods to nonlinear problems. However, this is a rather hard task due to the actual great gaps between both. Based on the Hadamard product approach, Chen et al. [2] derived two kinds of generalized matrix formulations in numerical approximation of nonlinear differential or integral operators. By using these unified formulations, this paper presents and verifies the simple relationship (theorem 3.1) between numerical analogue solutions of nonlinear differential operators and their Jacobian matrices. It is noted that theorem 3.1 is only applicable for an important special class of polynomial-only algebraic system of equations. However, in practice such polynomial-only systems have very widespread applications. The theorem paves a shortcut path to exploit the existing methods of solving linear problems to the complex polynomial-only nonlinear problems. Some significant results are immediately obtained by using theorem 3.1. First, so far there is not general and simple approach available for stability analysis in the numerical solution of nonlinear initial value problems. We here develop such a technique based on the application of theorem 3.1. Second, the linear iterative methods of algebraic equations such the SOR, Gauss-Seidel, and Jacobi methods were often applied in conjunction with the Newton method rather than directly to nonlinear system of equations itself. The present work directs to a straightforward extension of these techniques to nonlinear algebraic equations. Third, the quasi-Newton equation is the very basis of various quasi-Newton methods. Based on theorem 3.1, we constitute an exact alternative equation of the approximate quasi-Newton equation. As a consequence, we derive a set of the modified BFGS matrix updating formulas. Finally, in order to avoid the calculation of the Jacobian matrix and its inverse, we introduce the pseudo-Jacobian matrix. By using this new concept, the general nonlinear system of equations without limitation of polynomial-only problems is encompassed in this work. The proposed pseudo-Jacobian matrix is used for stability analysis of nonlinear initial value problems.

This paper is structured as follows. Section 2 gives a brief introduction to the Hadamard and SJT products. Two unified matrix formulations of general numerical discretization of nonlinear problems are obtained by using the Hadamard product. Section 3 proves the simple relationship theorem 3.1 between the numerical analogue of nonlinear operator and the corresponding Jacobian matrix, which forms the basis of later work. Section 4 is comprised of three subsections. Section 4.1 is concerned with stability analysis of numerical solution of nonlinear initial value problems, and in section 4.2 several existing linear iterative methods are directly extended to the nonlinear problems. Section 4.2 involves the construction of a set of the modified Broyden-type matrix updating formulas. Section 5 defines the pseudo-Jacobian matrix, and applies this concept to stability analysis of nonlinear initial value computation. Finally, some remarks of the present work are given in section 6. Unless otherwise specified, U, C and F in this paper represent vector.

## 2. Two unified matrix formulations of general nonlinear discretizations

Matrix computations are of central importance in nonlinear numerical analysis and computations. However, since nonlinear problems are actually different from linear ones, the traditional linear algebraic approach, which are based on the concept of linear transformation, can not provide a unified powerful tool for nonlinear numerical computation and analysis task. In this section, by using the Hadamard product and SJT product, we gain two kinds of generalized matrix formulations of nonlinear numerical discretization and a simple, accurate approach to calculate the Jacobian matrix [1, 2].

**Definition 2.1** Let matrices $A=[a_{ij}]$ and $B=[b_{ij}] \in C^{N \times M}$, the Hadamard product of matrices is defined as $A \circ B = [a_{ij} b_{ij}] \in C^{N \times M}$. where $C^{N \times M}$ denotes the set of N×M real matrices.

**Definition 2.2** If matrix $A=[a_{ij}] \in C^{N \times M}$, then $A^{\circ q}=[a_{ij}^q] \in C^{N \times M}$ is defined as the Hadamard power of matrix A, where q is a real number. Especially, if $a_{ij} \neq 0$, $A^{\circ(-1)}=[1/a_{ij}] \in C^{N \times M}$ is defined as the Hadamard inverse of matrix A. $A^{\circ 0}=11$ is defined as the Hadamard unit matrix in which all elements are equal to unity.

**Definition 2.3** If matrix $A=[a_{ij}] \in C^{N \times M}$, then the Hadamard matrix function $f^\circ(A)$ is defined as $f^\circ(A) = [f(a_{ij})] \in C^{N \times M}$.

**Theorem 2.1**: letting A, B and $Q \in C^{N \times M}$, then

1> $A \circ B = B \circ A$ (1a)

2> $k(A \circ B) = (kA) \circ B$, where k is a scalar. (1b)

3> $(A+B) \circ Q = A \circ Q + B \circ Q$ (1c)

4> $A \circ B = E_N^T (A \otimes B) E_M$, where matrix $E_N$ (or $E_M$) is defined as $E_N = [\,e_1 \otimes e_1 \vdots \cdots \vdots e_N \otimes e_N\,]$, $e_i = [0 \cdots 0 \underset{i}{1} 0 \cdots 0]$, i=1, ⋯, N, $E_N^T$ is the transpose matrix of $E_N$. ⊗ denotes the Kronecker product of matrices. (1d)

5> If A and B are non-negative, then $\lambda_{\min}(A) \min\{b_{ii}\} \leq \lambda_j(A \circ B) \leq \lambda_{\max}(A) \max\{b_{ii}\}$, where λ is the eigenvalue. (1e)

6> $(\det A)(\det B) \leq \det(A \circ B)$, where det( ) denotes the determinant. (1f)

For more details about the Hadamard product see [3, 4].

## 2.1. Nonlinear formulation-K[*] of general numerical methods

It is well known that the majority of popular numerical methods such as the finite element, boundary element, finite difference, Galerkin, least square, collocation and spectral methods have their root on the method of weighted residuals (MWR) [5, 6]. Therefore, it will be generally significant to apply the Hadamard product to the nonlinear computation of the MWR. In the MWR, the desired function $\phi$ in the differential governing equation

$$\psi\{u\} - f = 0 \qquad (2)$$

is replaced by a finite series approximation $\hat{u}$,

$$u = \hat{u} = \sum_{j=1}^{N} C_j \phi_j, \qquad (3)$$

where $\psi\{\ \}$ is a differential operator. $\phi_i$ can be defined as the assumed functions and $C_j$'s are the unknown parameters. The approximate function $\hat{u}$ is completely specified in terms of unknown parameters $C_j$. Substituting this approximation $\hat{u}$ into the governing equation (2), it is in general unlikely that the equation will be exactly satisfied, namely, result in a residual R

$$\psi\{\hat{u}\} - f = R \qquad (4)$$

The method of weighted residuals seeks to determine the N unknowns $C_j$ in such a way that the error R is minimized over the entire solution domain. This is accomplished by requiring that weighted average of the error vanishes over the solution domain. Choosing the weighting function $W_j$ and setting the integral of R to zero:

$$\int_D [\psi\{\hat{u}\} - f] W_j dD = \int_D R W_j dD = 0, \quad j=1,2,\ldots,N. \qquad (5)$$

Equation (5) can be used to obtain the N unknown coefficients. This equation also generally describes the method of weighted residuals. In order to expose our idea clearly, considering the following linear and nonlinear differential operators in two dimensions with varying parameter:

$$L_1\{u\} = c(x,y)\frac{\partial^p u}{\partial x^p} \qquad (6a)$$

$$L_2\{u\} = c(x,y)\frac{\partial^p u}{\partial x^p}\frac{\partial^q u}{\partial y^q} \qquad (6b)$$

Substitution of Eq. (3) into Eqs. (6a) and (6b) and applications of equation (1d) in the theorem 2.1 result in

$$L_1\{\hat{u}\} = c(x,y)\left\{\frac{\partial^p \phi_i}{\partial x^p}\right\}^T C, \tag{7a}$$

$$L_2(\hat{u}) = c(x,y)\left[\left\{\frac{\partial^p \phi_i}{\partial x^p}\right\}^T C\right] \circ \left[\left\{\frac{\partial^q \phi_i}{\partial y^q}\right\}^T C\right] = E_1^T c(x,y)$$

$$\left(\left\{\frac{\partial^p \phi_i}{\partial x^p}\right\}^T \otimes \left\{\frac{\partial^q \phi_i}{\partial y^q}\right\}^T\right) E_1(C \otimes C) = c(x,y)\left(\left\{\frac{\partial^p \phi_i}{\partial x^p}\right\}^T \otimes \left\{\frac{\partial^q \phi_i}{\partial y^q}\right\}^T\right)(C \otimes C) \tag{7b}$$

where $C$ is a vector composed of the unknown parameters, $E_1=1$. Substituting the Eqs. (7a, b) into Eq. (5), we have

$$\int_D L_1\{\hat{u}\} W_j dD = \left[\int_D c(x,y)\left\{\frac{\partial^p \phi_j}{\partial x^p}\right\}^T W_j dD\right] C \tag{8a}$$

$$\int_D L_2(\hat{u}) W_j dD = \left[\int_D c(x,y)\left(\left\{\frac{\partial^p \phi_i}{\partial x^p}\right\}^T \otimes \left\{\frac{\partial^q \phi_i}{\partial x^q}\right\}^T\right) W_j dD\right](C \otimes C) \tag{8b}$$

---
[*] It is denoted as formulation-S in [2].

As a general case, the quadratic nonlinear partial differential equation is given by

$$\sum_{k,l=0}^{N1} a_{kl}(x,y)\frac{\partial^{(k+l)}u}{\partial x^k \partial y^l} + \sum_{\substack{i,j=0 \\ k,l=0}}^{N2} b_{kl}(x,y)\frac{\partial^{(k+l)}u}{\partial x^k \partial y^l}\frac{\partial^{(i+j)}u}{\partial x^i \partial y^j} + d = 0, \tag{9}$$

where d is constant. The above equation encompasses a wide range of the quadratic nonlinear governing equations.

Applying Eqs. (8a, b), we can easily derive the MWR formulation of the nonlinear differential equation (9) with the following form

$$K_{n \times n} C + G_{n \times n^2}(C \otimes C) + F = 0, \tag{10}$$

where $F$ is the constant vector.

$$K_{n \times n} = \int_D \left(\sum_{k,l=0}^{N1} a_{kl}(x,y)\left\{\frac{\partial^{(k+l)}\phi_i}{\partial x^k \partial y^l}\right\}^T\right) W_j dD \in C^{n \times n}$$

and

$$G_{n \times n^2} = \int_D \left(\sum_{\substack{k,l=0 \\ i,j=0}}^{N2} b_{kl}(x,y)\left\{\frac{\partial^{(i+j)}\phi_p}{\partial x^i \partial y^j}\right\}^T \otimes \left\{\frac{\partial^{(k+l)}\phi_p}{\partial x^k \partial y^l}\right\}^T\right) W_j dD \in C^{n \times n^2}$$

represent constant coefficient matrices corresponding to the linear and nonlinear operators, respectively. For the cubic nonlinear differential equations, we can obtain similar general matrix formulation by using the same approach:

$$L_{n\times n}C + R_{n\times n^3}(C \otimes C \otimes C) + F = 0, \tag{11}$$

where L and R are constant coefficient matrices. For higher order nonlinear problems, the formulations can be easily obtained in the same way. To simplify notation, formulations with form of Eqs. (10) and (11) are denoted as formulation-K, where K is chosen since the Kronecker product is used in expressing nonlinear numerical discretization term.

As was mentioned earlier, most of popular numerical techniques can be derived from the method of weighted residual. The only difference among these numerical methods lies in the use of different weighting and basis functions in the MWR. From the foregoing deduction, it is noted that Eq. (10) can be obtained no matter what weighting and basis functions we use in the method of weighted residuals. Therefore, it is straightforward that we can obtain the formulation-K for the nonlinear computations of these methods. In many numerical methods, the physical values are usually used as the desired variables instead of the unknown expansion coefficient vector C in the preceding formulas. Both approaches are in fact identical. It is theoretically convenient to use C here.

In the following we give explicit formulas for computing the Jacobian derivative matrix of the quadratic and cubic nonlinear formulation-K (10) and (11). Eq. (10) can be restated

$$K_{n\times n}C + \begin{Bmatrix} vec(G_1) \\ \vdots \\ vec(G_n) \end{Bmatrix}(C \otimes C) + F = 0, \tag{12}$$

where vec( ) is the vector-function of a rectangular matrix formed by stacking the column of matrix into one long vector [7]. $G_i$'s are n×n symmetric matrices and can be easily obtained from the corresponding rows of the matrix G in Eq. (10) through the invert process of vec( ). Furthermore, we have

$$\left(K_{n\times n} + \begin{Bmatrix} C^T G_1 \\ \vdots \\ C^T G_n \end{Bmatrix}\right)C + F = 0, \tag{13}$$

where superscript T means the transpose of vector. According to the rule in differentiation of matrix function [4], the Jacobian derivative matrix of the above equation can be obtained by the following formula:

$$\frac{\partial \varphi\{C\}}{\partial C} = K_{n \times n} + 2 \begin{Bmatrix} C^T G_1 \\ \vdots \\ C^T G_n \end{Bmatrix}. \tag{14}$$

Similarly, the cubic nonlinear equation (11) can be restated as

$$\psi(C) = L_{n \times n} C + \begin{Bmatrix} \left[ R^1_{n \times n^2}(C \otimes C) \right]^T \\ \vdots \\ \left[ R^n_{n \times n^2}(C \otimes C) \right]^T \end{Bmatrix} C + F = 0. \tag{15}$$

The Jacobian matrix of the above equation can be evaluated by

$$\frac{\partial \psi\{C\}}{\partial C} = L_{n \times n} + \begin{Bmatrix} C^T \frac{\partial R^1_{n \times n^2}(C \otimes C)}{\partial \bar{C}} \\ \vdots \\ C^T \frac{\partial R^n_{n \times n^2}(C \otimes C)}{\partial C} \end{Bmatrix} + \begin{Bmatrix} \left[ R^1_{n \times n^2}(C \otimes C) \right]^T \\ \vdots \\ \left[ R^n_{n \times n^2}(C \otimes C) \right]^T \end{Bmatrix}. \tag{16}$$

Furthermore, we have

$$\frac{\partial \psi(C)}{\partial C} = L + 3 \begin{bmatrix} C^T R_{11} C & C^T R_{12} C & \cdots & C^T R_{1n} C \\ C^T R_{21} C & C^T R_{22} C & \cdots & C^T R_{2n} C \\ \vdots & \vdots & \vdots & \vdots \\ C^T R_{n1} C & C^T R_{n2} C & \cdots & C^T R_{nn} C \end{bmatrix}, \tag{17}$$

where $R_{ij}$ result from matrix $R^i$ and are rectangular constant coefficient matrices.

## 2.2. Nonlinear formulation-H and SJT product

The previously presented formulation-K is somewhat complex. This section will show that the Hadamard product can be directly exploited to express nonlinear discretization term of some numerical techniques. For example, consider the quadratic nonlinear differential operator $\frac{\partial U(x,y)}{\partial x} \frac{\partial U(x,y)}{\partial y}$, its numerical analogue by using a point-wise approximation technique can be expressed

$$\frac{\partial u(x,y)}{\partial x} \frac{\partial u(x,y)}{\partial y} = \{u_x u_y\}_i = \{u_x\}_i \circ \{u_y\}_i = \left( A_x \bar{U} \right) \circ \left( A_y \bar{U} \right), \quad i=1,2,\ldots,N, \tag{18}$$

where i indexes the number of discrete points; $A_x$ and $A_y$ represent the coefficient matrices dependent on specified numerical discretization scheme. It is noted that we use the desired function value vector $\bar{U}$ here instead of the unknown parameter vector C in section 2.1. In fact, both is equivalent. This explicit matrix formulation (18) is obtained in a straightforward and intuitive way. The finite difference, collocation methods and their variants such as differential quadrature and pseudo-spectral methods belong to the point-wise approximating numerical technique. In addition, the finite volume, dual reciprocity BEM [8] (the most efficient technique in applying BEM to nonlinear problems) and numerical techniques based on radial basis functions [9] can express their analogue of nonlinear differential operators in the Hadamard product form. For all above-mentioned methods, the numerical analogues of some nonlinear operators often encountered in practice are given by

1. $c(x,y)u_{,x} = \{c(x_j,y_j)\} \circ (A_x U)$,  (19a)

2. $(u_{,x})^q = (A_x U)^{\circ q}$, where q is a real number,  (19b)

3. $\dfrac{\partial u^m}{\partial x} \dfrac{\partial u^n}{\partial y} = (A_x U^{\circ m}) \circ (A_y U^{\circ n})$,  (19c)

4. $\sin u_{,x} = \sin^{\circ}(A_x U)$,  (19d)

5. $\exp(u_{,x}) = \exp^{\circ}(A_x U)$.  (19e)

In the above equations $(\ )_{,x} = \partial(\ )/\partial x$; $A_x$ and $A_y$ denote the known coefficient matrices resulting from cetain numerical methods. We define the nonlinear discretization expression in the Hadamard product form as the formulation-H. It is very easy to transform the formulation-H such as Eq. (18) into the formulation-K by using formula (1d). In what follows, the SJT product is introduced to efficiently compute analytical solution of the Jacobian derivative matrix.

**Definition 2.4.** If matrix $A=[a_{ij}] \in C^{N \times M}$, vector $U=\{u_j\} \in C^{N \times 1}$, then $A \lozenge U=[a_{ij} u_i] \in C^{N \times M}$ is defined as the postmultiplying SJT product of matrix A and vector U, where $\lozenge$ represents the SJT product. If M=1, $A \lozenge B = A \circ B$.

**Definition 2.5.** If matrix $A=[a_{ij}] \in C^{N \times M}$, vector $V=\{v_j\} \in C^{M \times 1}$, then $V^T \lozenge A = [a_{ij} v_j] \in C^{N \times M}$ is defined as the SJT premultiplying product of matrix A and vector V.

Considering the nonlinear formulation (18), we have

$$\dfrac{\partial}{\partial U}\{(A_x U) \circ (A_y U)\} = A_x \lozenge (A_y U) + A_y \lozenge (A_x U).$$  (20)

Formula (20) produces the accurate Jacobian matrix through simple algebraic computations. The SJT premultiplying product is related to the Jacobian matrix of the formulations such as

$$\frac{dU^m}{dx} = AU^m, \text{ i.e.,}$$

$$\frac{\partial}{\partial \bar{U}}\{A_x U^m\} = \left(mU^{\circ(m-1)}\right)^T \Diamond A_x. \tag{21}$$

In the following, we discuss some operation rules in applying the SJT product to evaluate the Jacobian matrix of the nonlinear formulations (13).

1. $\frac{\partial}{\partial \bar{U}}\left\{\{c(x_j, y_j)\} \circ (A_x U)\right\} = A_x \Diamond \{c(x_j, y_j)\}$ (22a)

2. $\frac{\partial}{\partial \bar{U}}\left\{(A_x U)^{\circ q}\right\} = qA_x \Diamond (A_x U)^{\circ(q-1)}.$ (22b)

3. $\frac{\partial}{\partial \bar{U}}\left\{(A_x U^{\circ m}) \circ (A_y U^{\circ n})\right\} = m\left\{\left(U^{\circ(m-1)}\right)\Diamond A_x\right\}\Diamond(A_y U^{\circ n}) +$
   $\qquad n\left\{\left(U^{\circ(n-1)}\right)\Diamond A_y\right\}\Diamond(A_x U^{\circ m})$ (22c)

4. $\frac{\partial}{\partial \bar{U}}\{\sin(A_x U)\} = A_x \Diamond \cos^\circ(A_x U)$ (22d)

5. $\frac{\partial}{\partial \bar{U}}\{\exp^\circ(A_x U)\} = A_x \Diamond \exp^\circ(A_x U)$ (22e)

6. If $\psi = f^\circ(\phi)$, $\phi = \varphi^\circ(U)$, we have $\frac{\partial \psi}{\partial U} = \frac{\partial \psi}{\partial \phi}\frac{\partial \phi}{\partial U}$. (22f)

In the above equations $\frac{\partial}{\partial \phi}$ and $\frac{\partial}{\partial U}$ represent the Jacobian derivative matrix of certain Hadamard vector function with respect to vectors $\phi$ and U, respectively. It is observed from these formulas that the Jacobian matrix of the nonlinear formulation-H can be calculated by using the SJT product in the chain rules similar to those in differentiation of a scalar function. The above computing formulas yield the analytical solutions of the Jacobian matrix. The computational effort of a SJT product is only $n^2$ scalar multiplications. However, it is noted that the SJT product seems to be not amenable to the evaluation of the Jacobian matrix of the previous formulation-K.

The finite difference method is often employed to calculate the approximate solution of the Jacobian matrix and also requires $O(n^2)$ scalar multiplications. In fact, the SJT product approach requires $n^2$ and $5n^2$ less multiplication operations than one and two order finite differences, respectively. Moreover, the SJT product produces the analytic solution of the Jacobian matrix. In contrast, the approximate Jacobian matrix yielded by the finite difference method affects the accuracy and convergence rate of the Newton-Raphson method, especially for highly nonlinear problems. The efficiency and accuracy of the SJT product approach were numerically demonstrated in [1, 2].

We notice the following fact that the SJT product is closely related with the ordinary product of matrices, namely, if matrix $A=[a_{ij}] \in C^{N \times M}$, vector $U=\{u_i\} \in C^{N \times 1}$, then the postmultiplying SJT product of matrix A and vector U satisfies

$$A \lozenge U = \text{diag}\{u_1, u_2, \ldots, u_N\} A, \tag{23a}$$

where matrix $\text{diag}\{u_1, u_2, \ldots, u_N\} \in C^{N \times N}$ is a diagonal matrix. Similarly, for the SJT premultiplying product, we have

$$V^T \lozenge A = A \, \text{diag}\{v_1, v_2, \ldots, v_M\}, \tag{23b}$$

where vector $V=\{v_j\} \in C^{M \times 1}$. The reason introducing the SJT product is to simplify the presentation, manifest the relation with the Jacobian matrix of the formulation-H and clear the fact that the SJT product approach enjoys the same chain rule of scalar differentiation.

Some numerical examples applying these above formulas presented in [1, 2]. The effectiveness and efficiency are demonstrated therein. Obviously, the formulation-H is preferred whenever possible. However, the formulation-K is believed to be in the most general an identical formulation form of various nonlinear numerical analogue due to its broad applicability. The general formulation-K and formulation-H provide a computational attractiveness to develop the unified techniques in the nonlinear analysis and computation. In next sections, we will employ the results given here.

## 3. Jacobian matrix and nonlinear numerical analogue

Consider the quadratic nonlinear term of equation (18) and its Jacobian matrix of equation (20), we have

$$\begin{aligned}
\left[A_x \lozenge (A_y U) + A_y \lozenge (A_x U)\right] U &= diag(A_y U)(A_x U) + diag(A_x U)(A_y U) \\
&= 2(A_x U) \circ (A_y U)
\end{aligned} \tag{24}$$

by means of equation (23a), where $diag(A_x U)$ and $diag(A_y U)$ are diagonal matrices with diagonal terms of $A_x U$ and $A_y U$. Furthermore, consider the cubic nonlinear differential operator

$$\frac{\partial u(x,y)}{\partial x} \frac{\partial u(x,y)}{\partial y} \frac{\partial^2 u(x,y)}{\partial xy} = (A_x U) \circ (A_x U) \circ (A_{xy} U), \tag{25}$$

whose Jacobian matrix is

$$A_x \Diamond [(A_y U) \circ (A_{xy} U)] + A_y \Diamond [(A_x U) \circ (A_{xy} U)] + A_{xy} \Diamond [(A_x U) \circ (A_y U)].$$

Similar to Eq. (24), we can derive

$$\{A_x \Diamond [(A_y U) \circ (A_{xy} U)] + A_y \Diamond [(A_x U) \circ (A_{xy} U)] + \\ A_{xy} \Diamond [(A_x U) \circ (A_y U)]\} U = 3(A_x U) \circ (A_y U) \circ (A_{xy} U) \tag{26}$$

In fact, we can summarize

$$N^{(2)}(U) = \frac{1}{2} J^{(2)}(U) U \tag{27}$$

for quadratic nonlinear term and

$$N^{(3)}(U) = \frac{1}{3} J^{(3)}(U) U \tag{28}$$

for cubic nonlinear term, where the $N^{(2)}$ and $N^{(3)}$ denote the quadratic and cubic nonlinear terms and $J^{(2)}$ and $J^{(3)}$ represent the corresponding Jacobian matrices.

As were mentioned in section 2, the formulation-K is in general appropriate for nonlinear numerical discretization expression of all numerical techniques resulting from the method of weighted residuals, which include the finite element, boundary element, finite difference, Galerkin, least square, collocation and spectral methods. Also, the nonlinear formulation-H of the finite difference, finite volume, dual reciprocity BEM, radial function based methods, collocation and their variants can easily be transformed into the formulation-K. Therefore, in the following we will investigate the effectiveness of equations (27) and (28) for the formulation-K. First, by comparing equations (13) and (14), it is very obvious

$$N^{(2)}(C) = \frac{1}{2} J^{(2)}(C) C \tag{29}$$

for the quadratic nonlinear formulation-K. Furthermore, by postmultiplying the Jacobian matrix of the cubic nonlinear term in formulation-K equation (17) by the vector C, we have

$$\begin{bmatrix} C^T R_{11} C & C^T R_{12} C & \cdots & C^T R_{1n} C \\ C^T R_{21} C & C^T R_{22} C & \cdots & C^T R_{nn} C \\ \vdots & \vdots & \vdots & \vdots \\ C^T R_{n1} C & C^T R_{n2} C & \cdots & C^T R_{nn} C \end{bmatrix} C = R_{n \times n^3} (C \otimes C \otimes C) \tag{30}$$

through inverse operations from equations (17) to (15). Therefore, we have

$$N^{(3)}(C) = \frac{1}{3} J^{(3)}(C) C \tag{31}$$

Next we use the mathematics inductive method to generalize the relationship formulas (29)

and (31) to any order nonlinear terms. First, we assume that there exists

$$N^{(k)}(C) = H^{(k)}_{n \times n^k} \left( \overbrace{C \otimes C \otimes \ldots \otimes C}^{k} \right) = \frac{1}{k} J^{(k)}(C) C \tag{32}$$

for the k-order nonlinear term. Consider the (k+1)-order nonlinear term, the corresponding formulation-K expression can be stated as

$$N^{(k+1)}(C) = H^{(k+1)}_{n \times n^{k+1}} \left( \overbrace{C \otimes C \otimes \ldots \otimes C}^{k+1} \right) = \begin{Bmatrix} \left[ N_1^{(k)}(C) \right]^T \\ \left[ N_2^{(k)}(C) \right]^T \\ \vdots \\ \left[ N_n^{(k)}(C) \right]^T \end{Bmatrix} C, \tag{33}$$

where $N_i^{(k)}(C)$ are the k-order nonlinear terms,

$$N_i^{(k)}(C) = h_{i_{n \times n^k}}^{(k+1)} \left( \overbrace{C \otimes C \otimes \ldots \otimes C}^{k} \right), \quad i=1, 2, \ldots, n. \tag{34}$$

The Jacobian matrix can be given by

$$J^{(k+1)}(C) = \frac{\partial N^{(k+1)}(C)}{\partial \bar{C}} = \begin{Bmatrix} C^T \dfrac{\partial N_1^{(k)}(C)}{\partial \bar{C}} \\ C^T \dfrac{\partial N_2^{(k)}(C)}{\partial \bar{C}} \\ \vdots \\ C^T \dfrac{\partial N_n^{(k)}(C)}{\partial \bar{C}} \end{Bmatrix} + \begin{Bmatrix} \left[ N_1^{(k)}(C) \right]^T \\ \left[ N_2^{(k)}(C) \right]^T \\ \vdots \\ \left[ N_n^{(k)}(C) \right]^T \end{Bmatrix} \tag{35}$$

By using equation (32), we have

$$\begin{aligned} J^{(k+1)}(C)C &= kN^{(k+1)}(C) + N^{(k+1)}(C) \\ &= (k+1)N^{(k+1)}(C) \end{aligned} \tag{36}$$

Therefore, it is generally validated

$$N^{(m)}(C) = \frac{1}{m} J^{(m)}(C) C, \tag{37}$$

where m denotes the nonlinear order. It is again stressed that the indirect parameter vector C formulation is actually equivalent to those using the unknown function value vector U. Therefore, equation (37) is equally satisfied for the vector U formulations. Summarize the above results, we have

**Theorem 3.1**: If $N^{(m)}(U)$ and $J^{(m)}(U)$ are defined as nonlinear numerical analogue of the m order nonlinear differential operator and its corresponding Jacobian matrix, respectively, then $N^{(m)}(U) = \frac{1}{m} J^{(m)}(U)U$ is always satisfied irrespective if which numerical technique is employed to discretize.

In fact, all integral-order nonlinear polynomial systems of equations can be represented in the formulation-K form. For example, consider quadratic nonlinear term $G_{n \times n^2}(C \otimes C)$ in equation (10), we can find that for an n-dimension polynomial system of equations, the quadratic nonlinear term can at most have $n^2$ independent coefficients. Therefore, coefficient matrix $G_{n \times n^2}$ is sufficient to determine any quadratic nonlinear polynomial terms uniquely. Similar analysis can be done for higher order nonlinear polynomial terms. Now we can conclude that theorem 3.1 is applicable for all integral-order nonlinear polynomial systems of equations. In addition, for quadratic nonlinear problems, $G_{n \times n^2}(C \otimes C)$ in equation (10) is a constant coefficient matrix and actually the second order derivative matrix (the Hessian matrix) of quadratic nonlinear algebraic vectors. Numerical properties such as singular values of such matrix may disclose some significant information of nonlinear systems.

## 4. Applications

By using theorem 3.1, this section will address some essential nonlinear computational issues pertinent to the computational stability analysis of nonlinear initial value problems, linear iteration solution of nonlinear algebraic equations and a modified BFGS quasi-Newton method.

### 4.1. Stability analysis of nonlinear initial value problems

The spatial discretization of time-dependent differential systems results in the initial value problem. For linear systems, methods for determining conditions of numerical stability and accuracy of various time integration schemes are well established. However, for nonlinear problems, these tasks have been dramatically complicated. It was found that numerical instability can occur in the nonlinear computation even for methods that are unconditionally stable for linear problems [10, 11]. Recently, an energy and momentum conserving condition is sometimes imposed to guarantee stability of nonlinear integration. The mathematical techniques in performing such strategy are often quite sophisticated and thus not easily learned and used. We wish to develop a methodology which can evaluate stability behavior of general integration schemes and avoids the above difficulties.

Without the loss of generality, the canonical form of the first order initial problem with nonlinear quadratic and cubic polynomial terms is given by

$$\frac{dU}{dt} = f(t,U)$$
$$= LU + N^{(2)}(t,U) + N^{(3)}(t,U) \tag{38}$$

where U is unknown vector; L is a given nonsingular matrix; $N^{(2)}(t, U)$ and $N^{(3)}(t, U)$ are given vectors of quadratic and cubic nonlinear polynomials, respectively. Therefore, according to theorem 3.1, we have

$$\left[L + \frac{1}{2}J^{(2)}(t,U) + \frac{1}{3}J^{(3)}(t,U)\right]U = A(t,U)U \tag{39}$$

where $J^{(2)}(t, U)$ and $J^{(3)}(t, U)$ are the Jacobian matrix of the quadratic and cubic nonlinear terms. It is seen that the right side of equation (38) is expressed as a definite explicit matrix-vector separated from in Eq. (39). So equation (38) can be restated as

$$\frac{dU}{dt} = A(t,U)U. \tag{40}$$

Eq. (40) has the form of linear initial value problem with varying coefficient matrix A(t, U), which provides a very attractive convenience to apply the well-developed techniques of linear problems to nonlinear problems.

A variety of linear time integration methods available now fall into two groups, explicit and implicit. The explicit methods are usually easy-to-use and need not solve a matrix system. However, these methods are also usually conditionally stable even for linear problems and thus in many cases stability requires small time step. For nonlinear problems, it is intrinsic advantages of the explicit methods that iterative solutions are not required. In wave propagation problems, the methods are often used due to their lower computing cost per step [12]. On the other hand, the implicit methods require the solution of a matrix system one or more times and therefore are computationally expensive per time step, especially for nonlinear problems. However, the implicit methods tend to be numerically stable and thus allow large time step. So these methods have advantage to solve stiff problems [13]. In what follows, we will investigate these two types of methods.

**Explicit methods**

As an example of the simplest, let us consider the explicit Euler scheme solution of equation

(38)

$$U_{n+1} = U_n + hf(t_n, U_n). \qquad (41)$$

In terms of equation (40), we have

$$U_{n+1} = [I + A(t_n, U_n)h]U_n, \qquad (42)$$

where I is unite matrix. If $A(t_n, U_n)$ is negative definite, then the stability condition is given by

$$h \prec \frac{2}{\lambda_{max}} \qquad (43)$$

as in the linear case, where $\lambda_{max}$ represents the largest eigenvalue of A. We note that $A(t_n, U_n)$ is a time-varying coefficient matrix. Therefore, it is difficult to predict stability behavior of global journey. In other words, the present strategy will only provide the local stability analysis at one time step. Park [10] pointed out that the local stable calculation can guarantee the global stability, inversely, the local instability causes the global response unstable. As in the linear system with time-varying coefficients, the key issue in the local stability analysis is how to determine $\lambda_{max}$. It is known [14] that the $l_p$ matrix norm of A gives a bound on all eigenvalues of A, namely

$$|\lambda_A| \le \|A\|. \qquad (44)$$

Of these, the $l_1$ or $l_\infty$ matrix norm of A is easily computed. Substitution of inequality (44) into inequality (43) produces

$$h \prec \frac{2}{\|A\|} \qquad (45)$$

Therefore, it is not difficult to confine the stepsize h satisfying stability condition inequality (43) by certain matrix norm.

For the other explicit integration methods, the procedure of stability analysis is similar to what we have done for the explicit Euler method. For example, the stable region of the negative real axis in the well-known fourth-order Runge--Kutta method is $|\lambda h| \prec 2.785$. Therefore, the method can be the local and global stable when applied to nonlinear problems provided that the condition

$$h \prec \frac{2.785}{\|A\|} \prec \frac{2.785}{\lambda_{max}} \qquad (46)$$

is satisfied.

In particular, we can obtain somewhat more elaborate results for the formulation-H given in section 2.2. To illustrate clearly, consider Burger's equation

$$\frac{\partial u}{\partial t} + u\frac{\partial u}{\partial x} = \frac{1}{Re}\frac{\partial^2 u}{\partial x^2} \qquad (47)$$

which Re is the Reynolds number. When applying the finite difference, collocation, finite volume method, dual reciprocity BEM or radial function based methods, the spatial discretization of the above equation will result in the formulation-H form discretization

$$\frac{dU}{dt} = \frac{1}{Re}B_x U - U \circ (A_x U), \qquad (48)$$

where U is consisted of the desired values of u. By using the SJT product, we get the Jacobian matrix of the right side nonlinear term of equation (48)

$$J^{(2)}(U) = -I \Diamond (A_x U) - A_x \Diamond U \qquad (49)$$

According to theorem 3.1, we have

$$\frac{dU}{dt} = A(t,U)U. \qquad (50)$$

where

$$A(t,U) = \frac{1}{Re}B_x - \frac{1}{2}\left[I \Diamond (A_x U) + A_x \Diamond U\right]. \qquad (51)$$

One can easily derive

$$\|A(t,U)\| \leq \frac{1}{Re}\|B_x\| + \|A_x\|\|U\|. \qquad (52)$$

Substituting inequality (52) into inequality (45), we have

$$h \leq \frac{2}{\frac{1}{Re}\|B_x\| + \|A_x\|\|U\|}. \qquad (53)$$

The above inequality gives the restriction conditions of stability when the explicit Euler method is applied to this case. If we have a priori knowledge of a bound of U, inequality (53) can provide global stability condition with respect to time stepsize h. For the fourth-order Runge-Kutta method, similar formulas can be obtained. It is seen from the preceding analysis that the present methodology of stability analysis deals with nonlinear problem in a similar way that we normally handle linear problems with time-dependent coefficients.

**Implicit and semi-implicit methods**

Without loss of generality, let us consider the implicit Euler method of the simplest implicit method as a case study

$$U_{n+1} = U_n + hf(t_n, U_{n+1}). \qquad (54)$$

In terms of equation (40), we have

$$U_{n+1} = [I - A(t_n, U_{n+1})h]^{-1} U_n. \tag{55}$$

As in the linear case, the stability condition of equation (55) is that coefficient matrix A(t, U) is negative definite. Due to the fact that $U_{n+1}$ is unknown before computation, the approach is a posterior stability analysis. In fact, for all A-stable implicit integration methods, the local and global stability can be guaranteed if the negative definite feature of matrix A is kept at all successive steps. It is noted that for A-stable integration methods, stability condition of nonlinear system is independent of the time step h as in linear case.

In the numerical solution of nonlinear initial value problems by implicit time-stepping methods, a system of nonlinear equations has to be solved each step in some iterative way. To avoid this troublesome and time-consuming task, various semi-implicit methods were developed by using linearization procedure in the implicit solution of nonlinear problem. For example, if the nonlinear function f(t, U) in the implicit Euler equation (54) is linearized by using Newton method, we get the semi-implicit Euler method, namely,

$$U_{n+1} = U_n + h\left[1 - h\frac{\partial f}{\partial U}\bigg|_{U_n}\right]^{-1} f(t, U_n), \tag{56}$$

where $\partial f / \partial U$ is the Jacobian matrix of f(t, U). The stability analysis of equation (56) can be carried out in the same way as we have done for the explicit methods.

**4.2. Linear iterative methods for nonlinear algebraic systems**

The linear iterative methods are used most often for large sparse system of linear equations, which include Jacobi, Gauss-Seidel and SOR methods. Newton method and its variants do not belong to this type of methods. Ortega and Rheinboldt [15] addressed the detailed discussions on various possible applications of these methods coupling the Newton-like methods to solve nonlinear problems such as the so-called SOR-Newton, Newton-SOR, etc. However, it is found very difficult when we attempt a direction extension of the linear iterative methods to nonlinear equations without the linearization procedure such as Newton method [14, p. 220, 15, p. 305]. This impediment stems from an explicit matrix-vector separated expression of the general nonlinear equations is not in general available. In this section, we confine our attention to overcome this barricade for the polynomial-only equations. Theorem 3.1 provides a simple approach to express the nonlinear terms as the explicit matrix-vector form. Therefore, it is possible to conveniently apply the general linear iterative methods to nonlinear polynomial-only systems, especially for the systems with the formulation-H form.

Without loss of generality, consider nonlinear polynomial equations

$$f(U) - b = 0 \qquad (57)$$

with the quadratic and cubic nonlinear terms. By using theorem 3.1, we have

$$\begin{aligned} f(U) &= \left[ L + \frac{1}{2} J^{(2)}(U) + \frac{1}{3} J^{(3)}(U) \right] U + b \\ &= A(U)U - b = 0 \end{aligned} \qquad (58)$$

where $J^{(2)}(U)$ and $J^{(3)}(U)$ are the Jacobian matrix of the quadratic and cubic nonlinear terms as in equation (39). Therefore, if we can easily compute $J^{(2)}(U)$ and $J^{(3)}(U)$. The obstacle in directly employing the linear iterative methods to nonlinear problems will be eliminated. By analogy with the original form of various linear iterative methods, we give the following nonlinear Jacobi, Gauss-Seidel and SOR iteration formulas in the solution of equation (58), respectively,

$$U_i^{(k+1)} = \frac{1}{a_{ii}\left(U_i^{(k)}\right)} \left( b_i - \sum_{j \neq i} a_{ij}\left(U_i^{(k)}\right) U_i^{(k)} \right), \qquad (59)$$

$$U_i^{(k+1)} = \frac{1}{a_{ii}\left(U_i^{(k)}\right)} \left( b_i - \sum_{j \prec i} a_{ij}\left(U_i^{(k)}\right) U_i^{(k+1)} - \sum_{j \succ i} a_{ij}\left(U_i^{(k)}\right) U_i^{(k)} \right), \qquad (60)$$

and

$$U_i^{(k+1)} = (1-\omega) U_i^{(k)} + \frac{\omega}{a_{ii}\left(U_i^{(k)}\right)} \left( b_i - \sum_{j \prec i} a_{ij}\left(U_i^{(k)}\right) U_i^{(k+1)} - \sum_{j \succ i} a_{ij}\left(U_i^{(k)}\right) U_i^{(k)} \right), \qquad (61)$$

where $\omega$ is the relaxation factor in the SOR method and allowed to vary with k. The choice of $\omega$ so as to maximize the rate of convergence of the SOR method is somewhat troublesome and seems to be problem-dependent, which has long attracted much attention. [15, 16] provide detailed discussions concerning $\omega$ for nonlinear cases. Ortega and Rheinboldt [15, p. 464-472] pointed out that the so-called M-function has the close relation with the nonlinear SOR method. According to theorem 3.1, it is obvious that for polynomial-only problems, the M-function is in fact dependent only on the Jacobian matrix of system. More work in this respect will be of vital importance. It is noted that no linearization procedure such as Newton method is used to get the above iterative formulas.

In particular, for nonlinear numerical analogue of the finite difference, finite volume, dual reciprocity BEM, radial function based methods, collocation and their variants, the discretization can be represented in the formulation-H form. We can use the SJT product method to yield the simple explicit matrix expression of the Jacobian matrix. Therefore, in fact, it is not necessary to evaluate the Jacobian matrix in these cases.

Of course, the initial start is also of considerable importance in applying these formulas. In addition, the multigrid method may be developed in coupling the Gauss-Seidel iterative formula (60) to accelerate convergence. The computational effort of the linear iterative methods is much less than the Newton method. However, as a penalty, the convergence rate is linear. It is noted that if $a_{ii}(U_i^{(k)})$ is found zero in iterative process, some row or column interchange is required. Some numerical experiments are also necessary to assess their performances vis-a-vis the Newton-like methods for various benchmark problems. Also, some in-depth results of linear iterative methods [14, 15] may be very useful to enhance the present nonlinear iterative formulas.

### 4.3. A modified BFGS quasi-Newton method

To avoid time-consuming evaluation and inversion of the Jacobian matrix in each iterative step of the standard Newton method, the quasi-Newton method was developed with maintaining a superlinear convergence rate. This key of such methods is a matrix-updating procedure, one of the most successful and widely used which is known as BFGS method named for its four developers, Broyden, Fletcher, Goldfarb, and Shanno. The so-called quasi-Newton equation is the very fundamental of various quasi-Newton methods, namely,

$$J_i(U_i - U_{i-1}) = f(U_i) - f(U_{i-1}). \tag{62}$$

The Jacobian matrix $J_i$ is updated by adding a rank-one matrix to the previous $J_{i-1}$ in satisfying equation (62) and the following relations:

$$J_i p = J_{i-1} p, \qquad \text{when } (U_i - U_{i-1})^T p = 0, \tag{63}$$

where $U_i - U_{i-1} = q$, $f(U_i) - f(U_{i-1}) = \delta f_i$. It is emphasized that J here is the Jacobian matrix of total system. It is noted that equation (62) is an approximate one. For the polynomial-only problems, we can gain the exact alternative of equation (62) by using theorem 3.1. Without loss of generality, equation (57) can be restated as

$$f(U) = LU + N^{(2)}(U) + N^{(3)}(U) + b = 0, \tag{64}$$

where LU, $N^{(2)}(U)$ and $N^{(3)}(U)$ represent the linear, quadratic and cubic terms of the system of equations. The Jacobian matrix of the system is given by

$$J = \frac{\partial f(U)}{\partial U} = L + \frac{\partial N^{(2)}(U)}{\partial U} + \frac{\partial N^{(3)}(U)}{\partial U}. \tag{65}$$

By using theorem 3.1, we have

$$JU = LU + 2N^{(2)}(U) + 3N^{(3)}(U) = \bar{f}(U). \tag{66}$$

Therefore, we can exactly establish

$$J_i U_i - J_{i-1} U_{i-1} = \bar{f}(U_i) - \bar{f}(U_{i-1}) = y \tag{67}$$

After some simple deductions, we get

$$J_i(U_i - U_{i-1}) = g, \tag{68}$$

where $g = -(J_i - J_{i-1})U_{i-1} + y$. It is worth stressing that equation (68) differs from equation (62) in that it is exactly constructed. In the same way of deriving the BFGS updating formula, applying equations (63) and (68) yields

$$J_i = J_{i-1} - \frac{(J_{i-1}q - g)q^T}{q^T q}. \tag{69}$$

Furthermore, we have

$$J_i \left( I + \frac{U_{i-1} q^T}{q^T q} \right) = J_{i-1} - \frac{(J_{i-1}q - J_{i-1}U_{i-1} - y)q^T}{q^T q}, \tag{70}$$

where I is the unite matrix. Note that left term in bracket of the above equation is the unite matrix plus a rank-one matrix. By using the known Shermann-Morrison formula, we can derive

$$J_i = J_{i-1} - \frac{(J_{i-1}U_{i-1})q^T}{q^T q + q^T U_{i-1}} - \frac{(J_{i-1}q - J_{i-1}U_{i-1} - y)q^T}{q^T q} \\ + \frac{(J_{i-1}q - J_{i-1}U_{i-1} - y)q^T(U_{i-1}q^T)}{(q^T q + q^T U_{i-1})q^T q} \tag{71}$$

The above equation (71) can be simplified as

$$J_i = J_{i-1} + rq^T \tag{72}$$

where

$$\gamma = -\frac{J_{i-1}U_{i-1}}{q^T q + q^T U_{i-1}} - \frac{(J_{i-1}q - J_{i-1}U_{i-1} - y)q^T}{q^T q} + \frac{(J_{i-1}q - J_{i-1}U_{i-1} - y)q^T U_{i-1}}{(q^T q + q^T U_{i-1})q^T q} \tag{73}$$

By employing the Shermann-Morrison formula to equation (72), we finally have

$$J_i^{-1} = J_{i-1}^{-1} - \frac{(J_{i-1}^{-1} r)(q^T J_{i-1}^{-1})}{1 + q^T (J_{i-1}^{-1} r)}. \tag{74}$$

The updating formulas (69) and (74) are a modified version of the following original BFGS formulas:

$$J_i = J_{i-1} - \frac{(J_{i-1}q - \delta f)q^T}{q^T q} \tag{75}$$

and

$$J_i^{-1} = J_{i-1}^{-1} - \frac{(J_{i-1}^{-1}\delta f - q)q^T J_{i-1}^{-1}}{q^T (J_{i-1}^{-1}\delta f)}, \tag{76}$$

where $\delta f$ is defined as in equation (63). One can find that modified updating formulas (69) and (74) look slightly more complicated compared to the original BFGS formulas (75) and (76), but in fact, the required multiplication number of both are nearly the same, only about $3n^2$ operations. In a similar way, equation (68) can also be utilized to derive the modified DFP quasi-Newton updating matrix formulas. Theoretically, the present updating formulas improve the accuracy of the solution by establishing themselves on the exact equation (68) instead of the approximate quasi-Newton equation (62). The basic idea of the quasi-Newton method is a successive update of rank one or two. Therefore, it is noted that equation (68) is not actually exact due to the approximate Jacobian matrix yielded in the previous iterative steps. It may be better to initialize Jacobian matrix J via an exact approach. In addition, it is a puzzle for a long time why one-rank BFGS updating formula performs much better compared to the other two-rank updating schemes such as DFP method [17]. In our understanding, the most possible culprit is due to the inexactness of quasi-Newton equation (62). Therefore, this suggests that the updating formulas of higher order may be more attractive in conjunction with equation (68), which will include more additional curvature information to accelerate convergence. It is noted that in one dimension, the present equations (69) and (74) degenerates into the original Newton method by comparing with the fact that the traditional quasi-Newton method becomes the secant method. The performances of the present methodology need be examined in solving various benchmark problems.

Also, the theorem 3.1 provides a practically significant approach to examine the deviation between the approximate and exact Jacobian matrices by vector norm

$$err[\hat{J}(U)] = \|\bar{f}(U) - \hat{J}(U)U\| / \|\bar{f}(U)\|, \tag{77}$$

where $\hat{J}(U)$ is the approximate Jacobian matrix of f(U), $f(U)$ and $\bar{f}(U)$ are defined in equations (64) and (66), respectively.

# 5. Pseudo-Jacobian matrix and its applications

The efficient numerical solution and analysis of nonlinear systems of algebraic equations usually requires repeated computation of the Jacobian matrix and its inversion. Function differences and hand-coding of derivatives are two conventional numerical methods for this task. However, the former suffers from possible inaccuracy, particularly if the problem is highly nonlinear. The latter method is time-consuming and error prone, and a new coding effort is also required whenever a function is altered. Recently, the automatic differentiation (AD) techniques receive an increased attention. However, straightforward application of AD software to large systems can bring about unacceptable amounts of computation. Either sparsity or structure of the systems is necessarily used to overcome the limitation. On the other hand, the SJT product approach presented in section 2 is a simple, accurate and efficient approach in the evaluation of the Jacobian matrix of the nonlinear systems with the formulation-H form. However, this approach is not applicable for general nonlinear system formulation-K. It is clear from the preceding review that a generally applicable, simple and efficient technique is, at least now, not available for the evaluation of the Jacobian matrix. In addition, the inversion of the Jacobian matrix is a more computationally expensive task. Our work in this section is concerned with the construction of the pseudo-Jacobian matrix of one rank to circumvent these difficulties. It is emphasized that the pseudo-Jacobian matrix presented below is in general applicable for any nonlinear algebraic systems with no restricted to polynomial-only problems.

Consider a general form of nonlinear initial value problem

$$\frac{dU}{dt} = LU + N(t,U), \tag{78}$$

where L is a given nonsingular matrix and N(t, U) is a given vector of nonlinear functions. N(t, U) can be expressed in a form

$$N(t,U) = \left\{ N(t,U) \frac{1}{n} \left( U^{\circ(-1)} \right)^T \right\} U$$
$$= \left( wv^T \right) U, \tag{79}$$

where n is the dimension size of equation system, $U^{\circ(-1)}$ is the Hadamard inverse of vector U as explained in definition 2.2 of section 2. It is necessary to bear in mind that all elements of unknown vector U can not be equal to zero when using formula (79). We can avoid the zero elements by using a simple linear transformation

$$\overline{U} = U + c, \tag{80}$$

where c is a constant vector. In the following discussion, we assume without loss of generality that no zero elements present in vector U. Therefore, by using formula (79), equation (78) can restated as

$$\begin{aligned}\frac{dU}{dt} &= \left[L + wv^T\right]U \\ &= A(t,U)U\end{aligned} \tag{81}$$

Note that w and v are vectors. Therefore, $wv^T$ is a matrix of one rank. Compared with equation (40), it is seen that both have an explicit expression with a separated matrix form. The difference lies in the fact that nonlinear terms are expressed as a one-rank modification to linear term in equation (81) with no use of Jacobian matrix, while equation (40) represents nonlinear terms via the respective Jacobian matrices. For convenient in notation, $wv^T$ is here defined as the pseudo-Jacobian matrix, which is a fundamental idea for the ensuing work. In fact, it is noted that the pseudo-Jacobian matrix can be derived for general linear and nonlinear terms with no limited applicable of polynomial-only systems.

As we have done in section 4.1, equation (81) can be applied to examine the local stability of the explicit and implicit methods when applied to nonlinear problems. For example, consider iterative equation (41) of the explicit Euler method and utilize stability condition inequalities (43) and (45), we have

$$h \prec \frac{2}{\left\|L + wv^T\right\|} \leq \frac{2}{\|L\| + \|w\|\|v^T\|}, \tag{82}$$

where w and v vary with $U_n$. Some elaborate results on $L + wv^T$ can be found in [18]. For the implicit method, consider iterative formula (54) of the back difference method, we have

$$U_{n+1} = \left[I - \left(L + wv^T\right)h\right]^{-1} U_n, \tag{83}$$

where w and v change with $U_{n+1}$. Therefore, for the A-stable back difference method, the local stability condition is to hold the negative definite of matrix $L + wv^T$.

It is emphasized here that the above procedure of applying the pseudo-Jacobian matrix to the stability analysis of the explicit Euler and implicit back difference methods is of equal applicability to all explicit, implicit and semi-implicit methods such as the Runge-Kutta, Rosenbrock, Gear backward difference and fully implicit Runge-Kutta methods, etc.

In the following, we establish the relationship between the pseudo-Jacobian matrix and original Jacobian matrix. Consider the nonlinear terms in equation (66), by using theorem 3.1, we have

$$J_N U = 2N^{(2)}(U) + 3N^{(3)}(U), \tag{84}$$

where $J_N$ represents the Jacobian matrix of the nonlinear terms. It is observed that equation (84) can not determine the Jacobian matrix $J_N$ uniquely in more than one dimension. By multiplying $\frac{1}{n}\left(U^{\circ(-1)}\right)^T$, we get

$$J_N U \left\{ \frac{1}{n}\left(U^{\circ(-1)}\right)^T \right\} = \left[ 2N^{(2)}(U) + 3N^{(3)}(U) \right] \left\{ \frac{1}{n}\left(U^{\circ(-1)}\right)^T \right\} \tag{85}$$
$$= \hat{J}_N$$

where $\hat{J}_N$ is the pseudo-Jacobian matrix of the orignal Jaobian matrix $J_N$. So we have

$$\hat{J}_N = J_N \frac{1}{n} \begin{bmatrix} 1 & \frac{U_1}{U_2} & \cdots & \frac{U_1}{U_n} \\ \frac{U_2}{U_1} & 1 & \cdots & \frac{U_2}{U_n} \\ \vdots & \vdots & \ddots & \vdots \\ \frac{U_n}{U_1} & \frac{U_n}{U_2} & \cdots & 1 \end{bmatrix} = J_N p(U), \tag{86}$$

where p(U) is defined as the deviation matrix between the original Jacobian and pseudo-Jacobian matrices for polynomial-only problems. Similar relationship for general nonlinear system is not available.

It is straightforward that this pseudo-Jacobian matrix can be employed to directly extend the linear iterative Jacobi, Gauss-Seidel and SOR methods to nonlinear system of equations. In addition, this concept can be applied to the Newton method to avoid the evaluation and inversion of Jacobian matrix. For the sake of brevity, they are not presented here.

## 6. Some remarks

In this study, the Jacobian matrix is established as a bridge between linear and nonlinear polynomial-only problems. Some significant results are achieved through the application of the theorem 3.1. It is worth stressing that although the theorem 3.1 was verified through the use of formulation-K and formulation-H given in section 2, it holds true no matter which approaches are employed in the expression of nonlinear analogue term and the evaluation of the Jacobian matrix. As was mentioned in section 2.1, any nonlinear algebraic polynomial-only equations can be expressed as the formulation-K form and theorem 3.1 can thus be exhibited for general nonlinear polynomial equations. For example, consider very simple mixed quadratic and cubic nonlinear algebraic equations [19]

$$\begin{cases} x_1^2 + x_2^2 - 1 = 0 \\ 0.75 x_1^3 - x_2 + 0.9 = 0 \end{cases} \tag{87}$$

It can be expressed in the formulation-K form as

$$\psi(x) = Lx + G_{2\times 4}(x \otimes x) + R_{2\times 8}(x \otimes x \otimes x) + F = 0, \tag{88}$$

and by using theorem 3.1, we have

$$\psi(x) = \left[ L + \frac{1}{2} J^{(2)}(x) + \frac{1}{3} J^{(3)}(x) \right] x + F = 0 \tag{89}$$

where $x = (x_1, x_2)$, $F$, $L$, $G_{2\times 4}$ and $R_{2\times 8}$ are the constant vector and coefficient matrices. $J^{(2)}(x)$ and $J^{(3)}(x)$ are the Jacobian matrix of the quadratic and cubic nonlinear terms, respectively. Nonlinear term XAX, in which X is a rectangular matrix of the desired values and A is constant coefficient matrix, often appears in optimal control, filter and estimation. Theorem 3.1 is the same effective for such nonlinear term. Interested readers may try more cases. It is interesting to note that equation (89) is in fact identical in form to the familiar derivative expression of scalar polynomial function. In the practical applications, it is not actually necessary to express nonlinear algebraic equations in the formulation-K form like equation (88). It is stressed that theorem 3.1 provide a convenient approach to express nonlinear system of equations as linear-form representation without the use of linearization procedure such as the Newton method. It is also well known that a very large class of real-world nonlinear problems can be modeled or numerically discretized polynomial-only algebraic system of equations. The results presented in this paper are in general applicable for all these problems. Therefore, this work is potentially important in a board spectrum of science and engineering. This paper is confined within the integral-order nonlinear problems. In fact, the theorem 3.1 is also applicable for fractional order nonlinear polynomial-only problems. We will further involve the problems of such type in the subsequent paper.

The concept of the pseudo-Jacobian matrix can be used for general nonlinear system of equations without restriction to polynomial-only problems. Due to its one-rank feature, the evaluation of inverse is avoided in various nonlinear computation and analysis, which results in a considerable saving in computing effort.

In sections 4.1 and 5, the explicit and implicit Euler methods of two simplest integrators are typically studied to avoid that the complexity of special integration methods obscures the exposition of the present fundamental strategy and make it hard to understand. It is very clear that the same procedure can be easily extended to nonlinear stability analysis of general explicit and implicit methods. For the A-stable methods, it is found that the local stability of solutions can be assured if the time-varying coefficient matrix sustains negative definite, which provides a clue how to construct some stable integrators for nonlinear initial value problems.

The present work may be in itself of theoretical importance and provides some innovative viewpoints for the nonlinear computation and analysis. Numerical examples assessing these given formulas and methodologies are presently underway. A further study of various possibilities applying the Hadamard product, SJT product, theorem 3.1 and pseudo-Jacobian matrix will be beneficial. For example, according to theorem 3.1, all nonlinear polynomial systems of equations can be expressed as a separated matrix linear system with variable-dependent coefficient matrix by using the Jacobian matrix. Therefore, the analysis of these Jacobian matrices may expose essential sources of some challenging problems such as numerical uncertainty [11], shock and chaotic behaviors.